\documentclass[journal=aamick,manuscript=article,layout=twocolumn]{achemso}

\usepackage{chemformula} 
\usepackage[T1]{fontenc} 
\usepackage{stfloats}

\author{Georgia Gkouzia}
\affiliation
{Institute of Materials Science, Technical University of Darmstadt, Germany}

\author{Damian Günzing}
\affiliation
{Faculty of Physics and Center for Nanointegration (CENIDE), University of Duisburg-Essen, Germany}
\alsoaffiliation
{Lawrence Berkeley National Laboratory, Berkeley, USA}

\author{Ruiwen Xie}
\affiliation
{Institute of Materials Science, Technical University of Darmstadt, Germany}

\author{Teresa Weßels}
\alsoaffiliation
{Faculty of Physics and Center for Nanointegration (CENIDE), University of Duisburg-Essen, Germany}
\affiliation
{Ernst Ruska-Centre for Microscopy and Spectroscopy with Electrons and Peter Gr\"unberg Institute, Forschungszentrum J\"ulich, Germany}

\author{Andr\'as Kov\'acs}
\affiliation
{Ernst Ruska-Centre for Microscopy and Spectroscopy with Electrons and Peter Gr\"unberg Institute, Forschungszentrum J\"ulich, Germany}

\author{Alpha T. N’ Diaye}
\affiliation
{Lawrence Berkeley National Laboratory, Berkeley, USA}

\author{Márton Major}
\affiliation
{Institute of Materials Science, Technical University of Darmstadt, Germany}

\author{J. P. Palakkal}
\alsoaffiliation
{Institute of Materials Science, Technical University of Darmstadt, Germany}
\affiliation
{Institute of Materials Physics, Georg-August-University of G\"ottingen, Germany}

\author{Rafal E. Dunin-Borkowski}
\affiliation
{Ernst Ruska-Centre for Microscopy and Spectroscopy with Electrons and Peter Gr\"unberg Institute, Forschungszentrum J\"ulich, Germany}

\author{Heiko Wende}
\affiliation
{Faculty of Physics and Center for Nanointegration (CENIDE), University of Duisburg-Essen, Germany}

\author{Hongbin Zhang}
\affiliation
{Institute of Materials Science, Technical University of Darmstadt, Germany}

\author{Katharina Ollefs}
\affiliation
{Faculty of Physics and Center for Nanointegration (CENIDE), University of Duisburg-Essen, Germany}

\author{Lambert Alff}
\affiliation
{Institute of Materials Science, Technical University of Darmstadt, Germany}

\email{georgia.gkouzia@tu-darmstadt.de, ruiwen.xie@tmm.tu-darmstadt.de}

\title{Element-Specific Study of Magnetic Anisotropy and Hardening in SmCo$_{5-x}$Cu$_{x}$ Thin Films}

\keywords{Permanent magnet, thin film, SmCo$_{5}$, magnetic hysteresis. }

\begin{document}


\begin{abstract}
This work investigates the effect of copper substitution on the magnetic properties of SmCo$_{5}$ thin films synthesized by molecular beam epitaxy. A series of thin films with varying concentrations of Cu were grown under otherwise identical conditions to disentangle structural and compositional effects on the magnetic behavior. The combined experimental and theoretical studies show that Cu substitution at the Co$_{3g}$ sites not only stabilizes the formation of the SmCo$_{5}$ structure but enhances magnetic anisotropy and coercivity. Density functional theory calculations indicate that Sm(Co$_4$Cu$_{3g}$)$_5$ possesses a higher single-ion anisotropy as compared to pure SmCo$_{5}$.
In addition, X-ray magnetic circular dichroism reveals that Cu substitution causes an increasing decoupling of the Sm 4\textit{f} and Co 3\textit{d} moments. Scanning transmission electron microscopy confirms predominantly SmCo$_{5}$ phase formation and reveals nanoscale inhomogeneities in the Cu and Co distribution. Our study based on thin film model systems and advanced characterization as well as modeling reveals novel aspects of the complex interplay of intrinsic and extrinsic contributions to magnetic hysteresis in rare earth-based magnets, \textit{i.e.} the combination of increased intrinsic anisotropy due to Cu substitution and the extrinsic effect of inhomogeneous elemental distribution of Cu and Co.  
\end{abstract}


\section{Introduction}
SmCo$_{5}$-based permanent magnets were developed in the 1960s and known to possess extremely strong uniaxial magnetic anisotropy of about $K_1=17.2$\, MJ/m$^3$ \cite{coey2010magnetism,strnat1967family,strnat1988rare}.  The large magnetocrystalline anisotropy energy (MAE) arises due to the spin-orbit coupling of localized and partially filled 4\textit{f} electrons of Sm and the spin-orbit coupling of the itinerant 3\textit{d} electrons of cobalt in a strong crystal electric field \cite{larson2003magnetic,ucar2020overview}. Besides, SmCo$_{5}$ exhibits a relatively large energy product ($BH$)$_{\text{max}}$ up to 200\,kJ/m$^3$ and a Curie temperature of 1020\,K. The SmCo$_{5}$ phase has a hexagonal crystal structure with a space group of $P6/mmm$. 

In practical applications, SmCo$_{5}$-based permanent magnets are widely used in motors, generators, actuators, etc \cite {gutfleisch2011magnetic}. These magnets are made from complex alloys, such as Sm-Co-Cu-Fe-Zr derived from the Sm$_{2}$Co$_{17}$-type, and require sophisticated heat treatments during fabrication, which lead to the emergence of a unique microstructure \cite{mishra1981microstructure,sepehri2017correlation}. The increased demand for high ($BH$)$_{\text{max}}$ magnets in particular for high-temperature applications in renewable energy technologies has spurred recent interest in the various phenomena leading to high coercivity  \cite{gutfleisch2011magnetic,gutfleisch2009high}. It is well-known that phase decomposition of the starting alloy results in a cellular microstructure of Sm$_{2}$Co$_{17}$ (2:17) with a SmCo$_{5}$ (1:5) intergranular phase, and a Zr-rich platelet \cite{maury1993genesis,duerrschnabel2017atomic}. The SmCo$_{5}$ phase is Cu-rich and the Sm$_{2}$Co$_{17}$ phase Fe-rich. Over the years, various studies have investigated the role of the individual elements Cu, Fe, and Zr in intrinsic and extrinsic magnetic properties\cite{staab2023hard}. However, a complete understanding that could help to overcome the so-called Brown’s paradox and help in the development of novel green magnets is still lacking \cite{hadjipanayis2000high,xiong2004microstructure,wu2020situ,shang2020effect}. 

Sm-Co thin films offer precise control over nanostructure, making them an ideal candidate for providing model structures to understand the specific role of individual defects on the magnetic properties. Nevertheless, due to the thermodynamic instability of the SmCo$_{5}$ phase, also in thin films complex decomposition phenomena do occur. We have recently discovered, a novel phase decomposition regime in molecular beam epitaxy (MBE) grown thin films, resulting in the coexistence of SmCo$_{5}$ and Sm$_{2}$Co$_{17}$ blocks at the nanoscale with a width of only a few nanometers\cite{sharma2021epitaxy}. These films have low coercivity due to their high crystallinity, phase purity, and fully coherent interfaces between SmCo$_{5}$ and Sm$_{2}$Co$_{17}$. In contrast, sputtered thin films grown at much higher particle energies have a more complex precipitation nanostructure made up of various Sm-rich phases such as SmCo$_{3}$ and Sm$_{2}$Co$_{7}$, which can lead to ultra-high coercivity primarily due to pinning at low-symmetry grain boundaries\cite{akdogan2015preparation}. 

Typically, different buffers and underlayers made from Ru, Cr, or Cu have been used to ease the growth of Sm-Co thin films. Textured Cr and Ru buffer layers promote the $c$-axis orientation and improve the magnetic properties. Above all, Cu has been widely studied because it not only facilitates the out-of-plane orientation but has been found to have the unique advantage of stabilizing the CaCu$_{5}$ structure \cite{seifert2009epitaxial,hofer1970physical,zhang2003effect}. Nevertheless, the magnetic properties of the Sm-Co layer will be affected by diffusion processes from these buffer layers.  

In this work, we have established a direct growth process of Cu substituted SmCo$_{1-x}$Cu$_{x}$ samples on sapphire without using any additional underlayer, which eliminates the potential impact of diffusing elements. We use computational and experimental methods to characterize the element-specific magnetic properties and illuminate the role of Cu incorporation in SmCo$_5$ and its effect on coercivity. 

\section{Methodology}
\subsection{Experimental part}

The base pressure of the used MBE chamber was 10$^{-10}$\,mbar. The SmCo$_{5-x}$Cu$_{x}$ thin films were deposited by e-beam co-evaporation from elemental Sm, Co, and Cu sources. The films were deposited onto $c$-axis oriented Al$_{2}$O$_{3}$ substrates, known to promote the $c$-axis orientation of the SmCo$_{5}$ phase. First, a temperature scan was carried out showing that the most favorable temperature for growing a crystalline SmCo$_{5}$ layer was in this setup 550$^\circ$C. The substrates were heated from the backside using a diode laser, a technique that avoids contamination from in vacuum heaters. Before evaporation, the substrates were annealed at 540$^\circ$C for 1h in the MBE chamber, to obtain a clean crystalline substrate surface. For all samples, the deposition rate of the samarium was kept constant at 0.1\,{\AA}/sec. The individual deposition rates of cobalt and copper were changed but the sum of the deposition rate (Co plus Cu) was kept constant at 0.1\,{\AA}/sec. A series of SmCo$_{5-x}$Cu$_{x}$ films was produced with $x=0.5,1,1.5,2$. The evaporation rates during the growth of the thin films were controlled by using quartz crystal microbalances. The quality of the films was monitored using in-situ reflection high energy electron diffraction (RHEED).

X-ray diffraction (XRD) with Cu K$_{\alpha}$ radiation on a Rigaku SmartLab system was carried out for the crystallographic and structural characterization of the films. The thicknesses of the films were determined to be 30\,nm$\pm$2\,nm. The magnetic properties of the films were measured by a superconducting quantum interference device (SQUID) in two directions, out-of-plane and in-plane of the substrate surface, using the MPMS XL magnetometer by Quantum Design. The measurements have been performed at 300 K with external fields up to 6 Tesla. The diamagnetic contribution from the Al$_{2}$O$_{3}$ substrate has been subtracted by correcting the slope between 4 and 6 Tesla. The coupling of Sm and Co moments has been studied using X-ray magnetic circular dichroism (XMCD) by recording element-specific hysteresis loops. For structural and spectroscopic characterizations, scanning transmission electron microscopy (STEM) and energy dispersive X-ray spectroscopy (EDX) \cite{Chemi2016} was used in combination on cross-sectional specimens that were prepared along the $c$-axis using a focused Ga ion beam (FIB) sputtering. The STEM images and EDX maps were processed using Velox software (ThermoFisher Scientific). 

\subsection{Computational part} \label{comp}
In order to evaluate the single-ion anisotropy of Sm$^{3+}$, the crystal field parameters (CFPs) were calculated in the framework of density functional theory (DFT) using the WIEN2k program~\cite{blaha2020wien2k}. The generalized gradient approximation (GGA) was employed for the exchange-correlation functional. The experimental lattice parameters of SmCo$_{5}$ were adopted in the calculations. Regarding the Cu-doped cases, including Sm(Co$_{4}$Cu$_{3g}$)$_{5}$ and Sm(Co$_{4}$Cu$_{2c}$)$_{5}$, the lattice parameters were fixed to those of SmCo$_{5}$ considering the relatively small volume change. Therefore, the solely chemical effect of Cu can be explicitly probed. The $RMT \times K_{\text{max}}$ was set to 7 and a $k$-mesh of $9 \times 9 \times 9$ was sampled in the Brillouin zone. For the calculation of CFPs, we followed the method proposed by Nov{\'a}k \textit{et} \textit{al.}~\cite{novak2013crystal}, in which the local Hamiltonian in the basis of Wannier functions is expanded by a series of spherical tensor operators. In specific, the self-consistent field (SCF) calculation was first performed without spin polarization and with 4\textit{f} electrons in the core. Subsequently, a non-SCF calculation was carried out treating 4\textit{f} as valence states so that the 4\textit{f} states were allowed to hybridize with the transition metal 3\textit{d} states. In addition, we shifted the energy of 3\textit{d} states 0.4 Ry lower to assure appropriate hybridization strength. The Bloch states from the 4\textit{f} energy window were then transformed to Wannier functions using the wien2wannier interface~\cite{kunevs2010wien2wannier} followed by standard Wannierization process by Wannier90~\cite{mostofi2008wannier90}.  

The obtained CFPs were then used to construct the atomic Hamiltonian of Sm by including also the Coulomb interactions ($\hat{H}_{U}$), the spin-orbit coupling and the Sm-transition metal exchange coupling ($\hat{H}_{ex}$)
\begin{equation}
    \hat{H}_{at} = \hat{H}_{U} + \lambda \sum_i \hat{s}_i \hat{l}_i + \hat{H}_{CF} + \hat{H}_{ex}.
\end{equation}
The eigenvalue of the Hamiltonian was solved using the Lanczos algorithm as implemented in Quanty code~\cite{lu2014efficient}. By varying the magnetization direction corresponding to the exchange coupling term, the eigenvalue was then obtained as a function of azimuthal angle, which give rise to the single-ion anisotropy.

\section{Results and Discussion}

\subsection{Crystal Structure}

\begin{figure*}
    \centering
    \includegraphics[width=\linewidth]{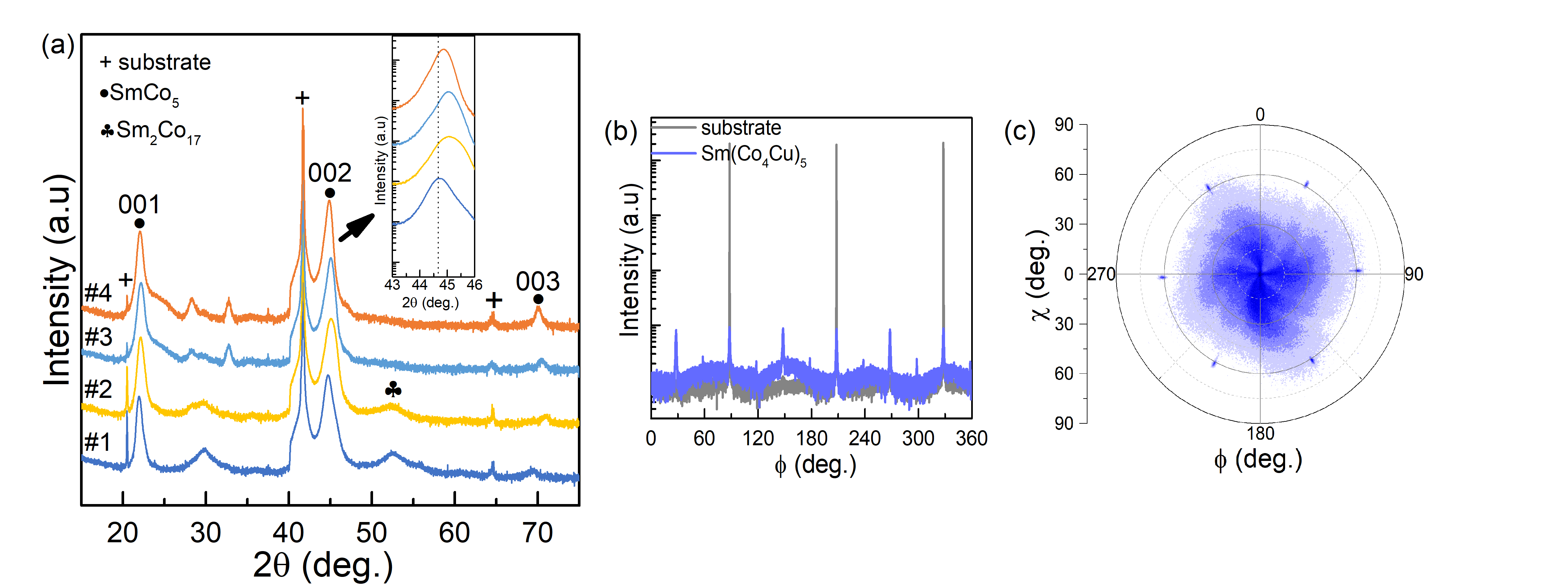}
    \caption{(a) $\theta$-2$\theta$ X-ray diffraction patterns of ($\#1$) SmCo$_{4.5}$Cu$_{0.5}$, ($\#2$) SmCo$_{4}$Cu, ($\#3$) SmCo$_{3.5}$Cu$_{1.5}$ and ($\#4$) SmCo$_{3}$Cu$_{2}$ thin films grown onto single crystalline Al$_{2}$O$_{3}$ substrate at 550$^\circ$C. At 41.67$^\circ$ the 006 reflection of the Al$_{2}$O$_{3}$ substrate is marked with the plus sign.
    (b) $\phi$ scan of the SmCo$_{4}$Cu film (blue) grown on single crystalline Al$_{2}$O$_{3}$ substrate (grey) at 550$^\circ$C.(c) Pole figure of the SmCo$_{5}$ reflection of the SmCo$_{4}$Cu film.}
    \label{fig:XRD}
\end{figure*}

Figure \ref{fig:XRD}a shows the XRD patterns of the SmCo$_{5-x}$Cu$_{x}$ films grown onto 001-oriented Al$_{2}$O$_{3}$ substrates. The main 006 reflection of the substrate appears at 41.67$^\circ$. The 001-oriented Al$_{2}$O$_{3}$ promotes the growth of $c$-axis textured Sm(Co,Cu)$_{5}$ films, indicated by the presence of \textit{00l}-type reflections. In the film with the lowest Cu content ($\#1$), the main reflection 002 of the SmCo$_{5}$ phase is observed at 44.218$^\circ$. Upon increasing the Cu content, the 002 reflection shifts to 45.106$^\circ$ and with an even further increase of Cu content ($\#4$), the reflection shifts again to lower angles (inset Figure \ref{fig:XRD}a). At the same time, the residual reflection of the Sm$_{2}$Co$_{17}$ phase decreases and eventually disappears upon increasing Cu content. The observed non-linear peak shift can be consistently explained as follows: First, the reduction of the 2:17 phase shifts the peak towards the 1:5 peak, as observed previously \cite{sharma2021epitaxy}. Then, the increasing Cu content shifts the peak back, due to the slightly enlarged lattice constant of SmCu$_{5}$, in agreement with Vegard's law. Note that the phase purity of the  1:5 phase is significantly higher than that of typical sputtered thin films, however, also reflections from a small amount of residual Sm-rich phases can still be detected around 29$^\circ$ and 30$^\circ$. Nevertheless, the overall microstructure remains in a consistent way predominantly 1:5 throughout the series.

Azimuth scans have been used to confirm crystal lattice symmetry and epitaxial relations. Here, we show a $\phi$-scan of the diffraction peak of the SmCo$_{4}$Cu sample with respect to the substrate. Figure \ref{fig:XRD}b shows the 104 reflection (grey 2$\theta$=35.03$^\circ$ and $\chi$=38.02$^\circ$) of the substrate which is rhombohedral and the SmCo$_{4}$Cu sample (blue) which is hexagonal. The $\phi$-scan shows three peaks which indicate the thee-fold symmetry of the Al$_{2}$O$_{3}$ substrate. Six peaks are obtained for the SmCo$_{4}$Cu film which shows the six-fold symmetry and proves its hexagonal phase. The observed peaks correspond to the 201 reflection of the SmCo$_{5}$ phase. The pole figure of the 201 reflections, Figure \ref{fig:XRD}c, shows the high-intensity Bragg peaks in blue. The distribution of the reflections indicates a crystalline, highly textured film.

\subsection{Magnetization measurements}

\begin{figure}
    \centering
    \includegraphics[width=\linewidth]{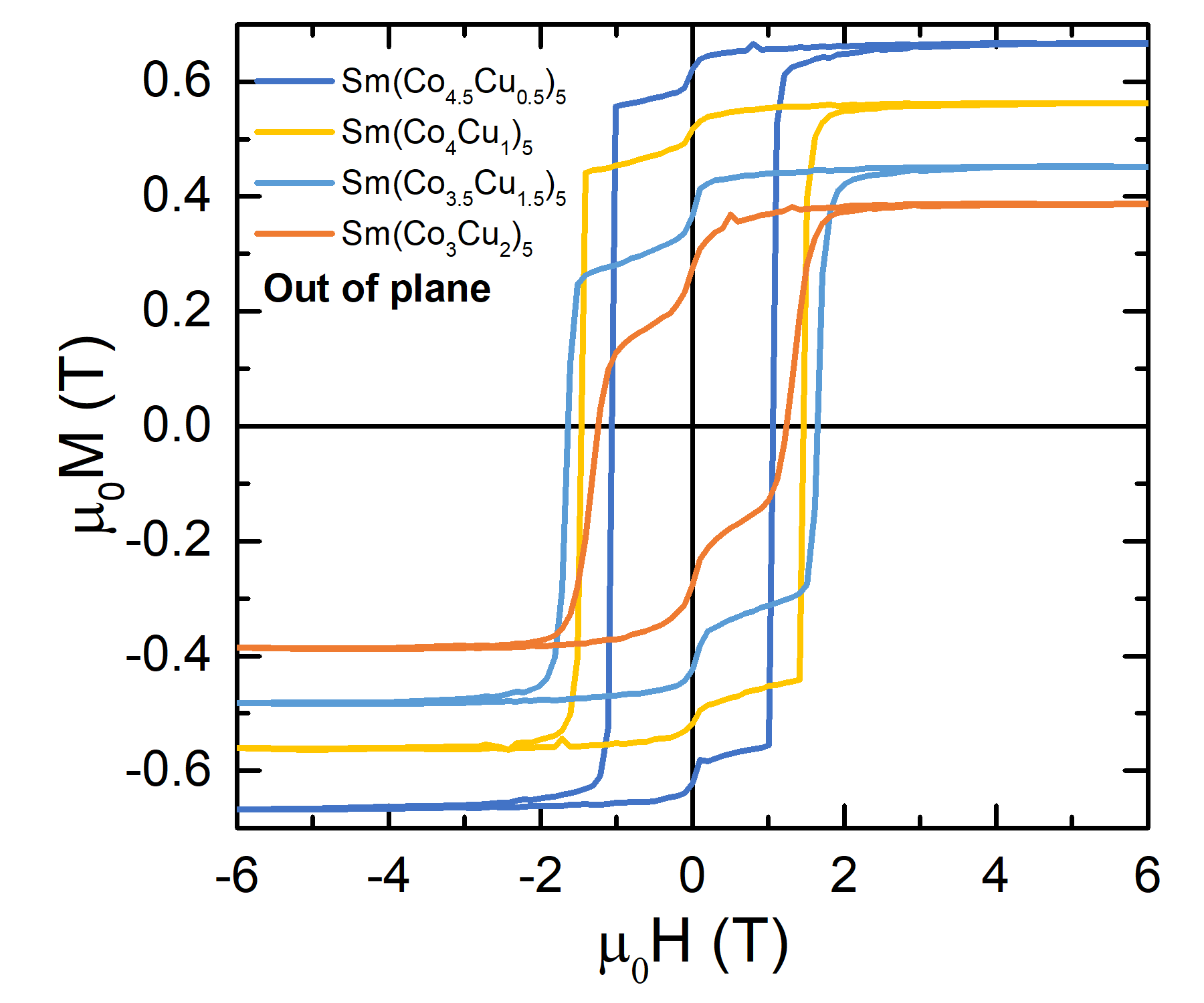}
    \caption{Magnetization curves of the SmCo$_{5-x}$Cu$_{x}$ thin films measured in the out-of-plane (OOP) direction as a function of applied field at 300 K.}
    \label{fig:OOP}
\end{figure}

\begin{figure}
    \centering
    \includegraphics[width=\linewidth]{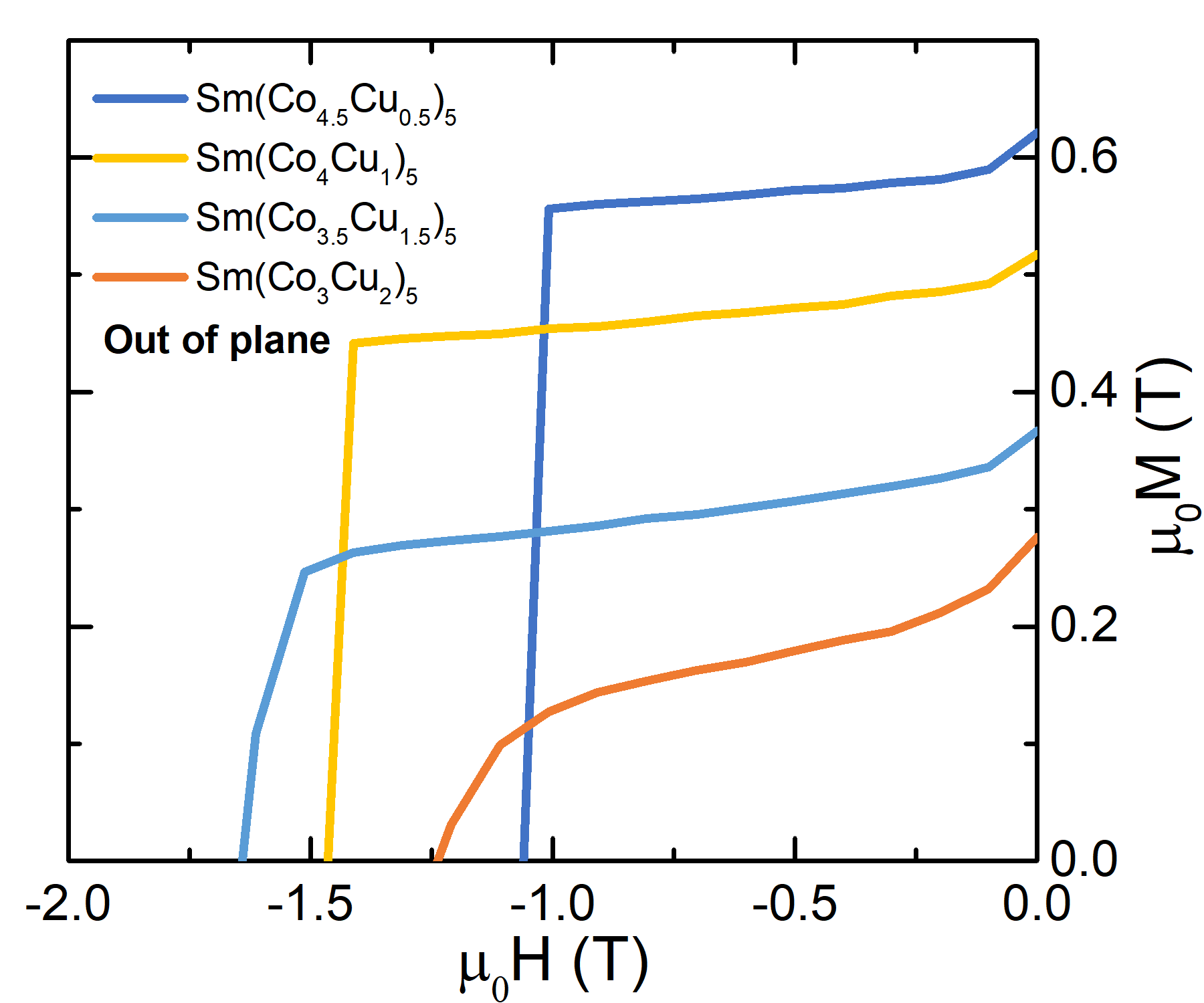}
    \caption{Demagnetization curves of the SmCo$_{5-x}$Cu$_{x}$ thin films as a function of applied field at 300 K.}
    \label{fig:Hc}
\end{figure}

\begin{figure}
    \centering
    \includegraphics[width=\linewidth]{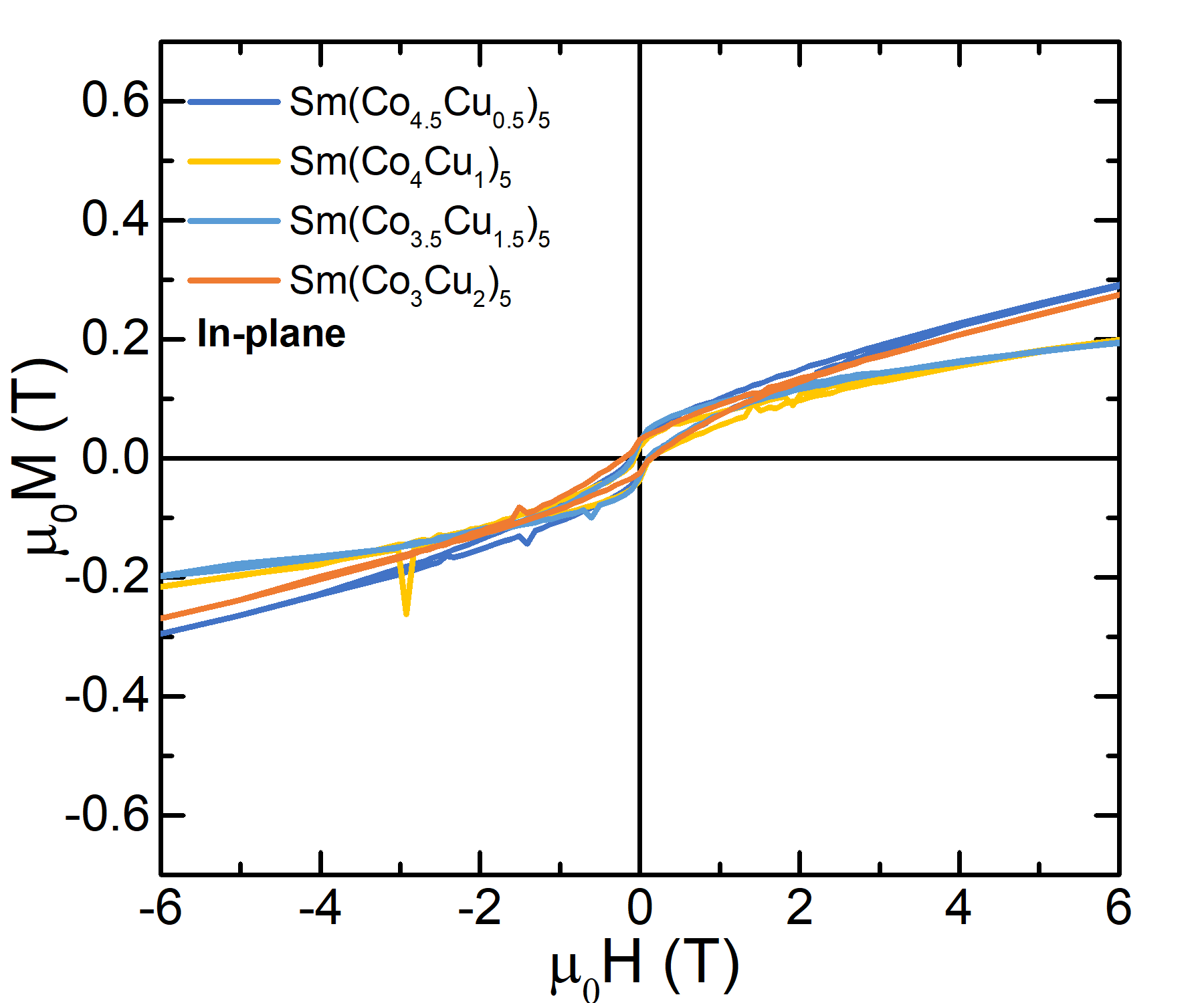}
    \caption{Magnetization curves of the SmCo$_{5-x}$Cu$_{x}$ thin films measured in the in-plane (IP) direction as a function of applied field at 300 K. }
    \label{fig:IP}
\end{figure}

Cu is usually used as dopant in SmCo$_{5}$ thin films grown onto Cr or Ru buffer layers \cite{yin2013effects,cui2016influence}. As mentioned above, in this work, no additional underlayers have been used, and the SmCo$_{5-x}$Cu$_{x}$ films were deposited directly on top of Al$_{2}$O$_{3}$ substrates. The hysteresis loops of the out-of-plane direction are shown in Figure \ref{fig:OOP}. The easy axis of magnetization is out of plane for all samples while the hard axis is in-plane as shown in Figure \ref{fig:IP}. Starting from the film with the lowest Cu concentration, a remanent magnetization of 0.6 T can be observed and the coercivity reaches 1.08 T. Upon increasing the Cu ratio, the coercivity drastically increases up to 1.64 T whereas the remanent and saturation magnetization, as expected due to Cu dilution of Co moments, decreases (see Fig.~\ref{fig:Hc}). A further increasing of Cu results in a rapid decrease of coercivity to 1.23 T. 

There are two sites where Cu can substitute Co, namely Co$_{2c}$ and Co$_{3g}$.\cite{haider2021enhancement} As the thin film microstructure is similar in all cases predominantly 1:5, we assume that extrinsic contributions to the coercivity are comparable - but with the exception of the distribution of the substitutional element copper. The Cu distribution will be addressed in section~\ref{TEM}. Therefore, we suggest that the increased coercivity is at least partly correlated with an increased intrinsic anisotropy resulting from Cu substitution at the Co$_{3g}$ sites. A similar suggestions has been made for Y(Co,Cu)$_{5}$ \cite{okumura2022first}, a system which also behaves in thin films similar to the here investigated Sm(Co,Cu)$_{5}$ \cite{sharma2021epitaxy,sharma2020y-co}.  
To corroborate this hypothesis, we discuss the described DFT-based modelling in section~\ref{discussion}. The in-plane magnetization measurements shown in Figure \ref{fig:IP} further support an increased anisotropy for SmCo$_{4}$Cu and SmCo$_{3.5}$Cu$_{1.5}$. Another factor leading to increased intrinsic anisotropy, is the improved crystallization of the 1:5 phase due to Cu substitution \cite{ohtake2010effects}. Interestingly, in the out-of-plane loops shown in Figure \ref{fig:Hc}, a small kink appears at zero field. This kink might be associated with the presence of a residual soft magnetic nanocrystalline or amorphous phase that is not strongly coupled to the hard magnetic phase\cite{kneller1991exchange}. 

\subsection{Element-specific measurements}
X-ray magnetic circular dichroism (XMCD) spectra and element-specific hysteresis curves have been recorded at the bending magnet beamline 6.3.1 of the Advanced Light Source (ALS) at Lawrence Berkeley National Laboratory. The measurements were carried out using an applied magnetic field of up to 1.9 T with a fixed polarization, at ambient temperatures. The samples have been measured in out-of-plane direction, parallel to the magnetic easy $c$-axis. As a detection method X-ray excited optical luminescence (XEOL) was used, which probes the full thickness of the film, due to the luminescence of oxygen in the Al$_{2}$O$_{3}$ substrate\cite{kallmayer2007interface}. In this way element-specific hysteresis loops have been recorded at fixed energies at the XMCD maxima and minima for Co L$_{2,3}$-edges (779.2 and 794.5~eV) and Sm M$_{4,5}$-edges (1082.5 and 1109.3~eV). 

\begin{figure}
    \centering
    \includegraphics[width=\linewidth]{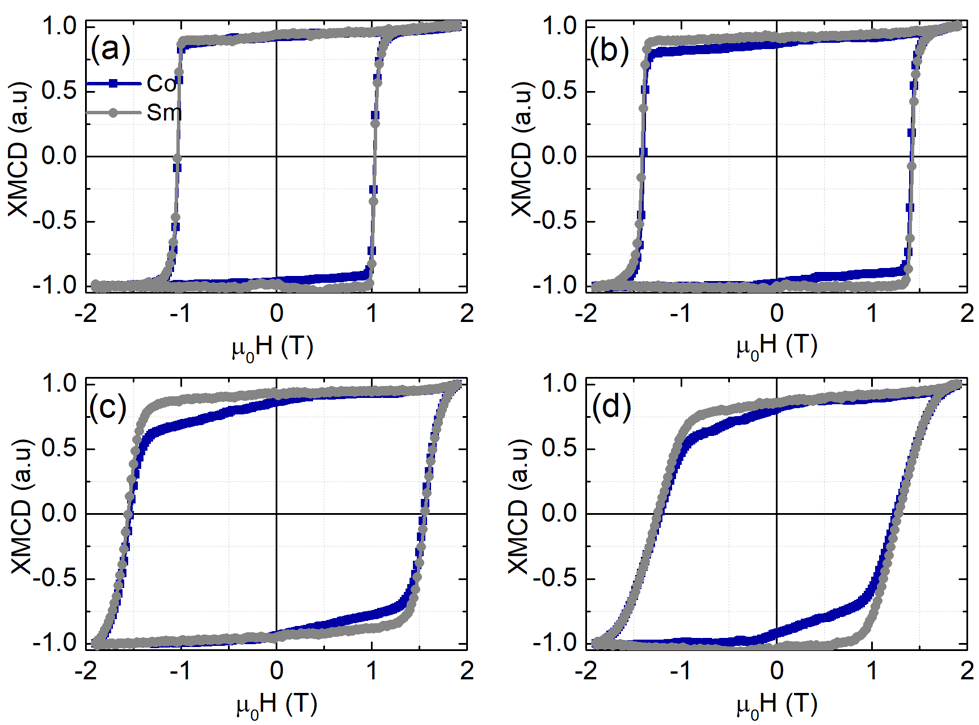}
    \caption{Element specific hysteresis loops for the Sm M$_{4,5}$ (grey) and Co L$_{2,3}$-edges (blue) recorded at room temperature applying positive and negative 1.9 T external fields in out-of-plane direction for the following samples (a) SmCo$_{4.5}$Cu$_{0.5}$ (b) SmCo$_{4}$Cu (c) SmCo$_{3.5}$Cu$_{1.5}$ and (d) SmCo$_{3}$Cu$_{2}$.}
    \label{fig:element-specific}
\end{figure}

The strength of element-specific hysteresis measurements is the quantitative separation of the individual contributions from the Co and Sm magnetic moment \cite{stohr1999exploring}. The normalized element-specific hysteresis curves are shown in Figure \ref{fig:element-specific}. The Co and Sm curves do not deviate significantly for the SmCo$_{4.5}$Cu$_{0.5}$ film shown in Figure \ref{fig:element-specific}a. This confirms the stronger exchange coupling between the rare earth Sm and the transition metal Co. Upon the increase of Cu substitution, a gap between Co and Sm appears in the second quadrant's demagnetizing curves. The measured data indicates that the substitution of Cu in the Co sub-lattice causes a continuous decoupling of the Sm and Co moments and softens the Co moment. The dilution with Cu is also the reason for the reduced total magnetization of the films. In contrast, samarium moments are only softened at higher Cu concentrations, also supported by the reduced critical temperature of the Cu-rich compounds.

\subsection{Structure and chemical composition of SmCo$_{4}$Cu}\label{TEM}

\begin{figure}[h!]
    \centering
    \includegraphics[width=\linewidth]{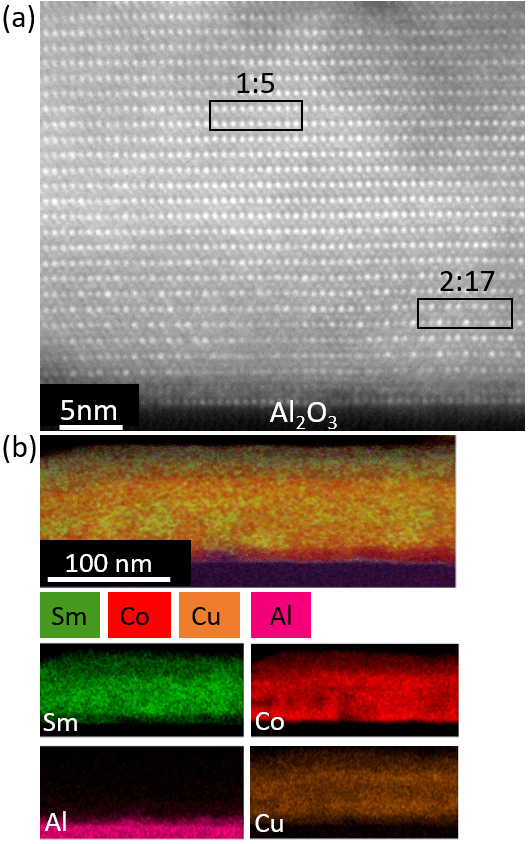}
    \caption{(a) High resolution HAADF STEM image showing the interface of the SmCo$_{4}$Cu thin film to the Al$_{2}$O$_{3}$ substrate. The dumbbell pattern of atomic columns reveals the presence of the Sm$_{2}$Co$_{17}$ phase. 
    (b) Chemical composition mapping of Sm, Co and Cu extracted from STEM/EDX spectrum imaging. Note the  distribution of Co and Cu is not uniform.}
    \label{fig:HAADF}
\end{figure}

The structure and elemental distribution of the SmCo$_{4}$Cu sample was studied in cross-sectional geometry. Figure \ref{fig:HAADF}(a) shows the structure of the sample in a high resolution high-angle annular dark-field (HAADF) STEM image. Note the presence of a characteristic "dumbbell" pattern in the Sm-Co-Cu that indicates the formation of the 2:17 phase at the interface to the Al$_2$O$_3$ substrate. The rest of the layer shows a uniform 1:5 phase formation. The thickness of the 2:17 phase is approximately 7\,nm. The transition from 2:17 to 1:5 phase is continuous \textit{i.e.}, no sharp boundaries were observed. 

Figure \ref{fig:HAADF}(b) shows the chemical composition distribution of individual elements extracted from EDX/STEM spectrum imaging. The Co and Cu maps reveal inhomogenous distribution of the elements suggesting Cu- and Co-rich regions. These inhomogeneities might be responsible for the zero field kink in the SQUID measurement which is absent in the XMCD measurement (see  Figs.~\ref{fig:Hc} and Fig.~\ref{fig:element-specific}). 

\subsection{Discussion}\label{discussion}

The magnetization curves of the SmCo$_{5-x}$Cu$_{x}$ films measured at room temperature were shown in Figure \ref{fig:OOP}. It is clear that Cu substitution affects the coercivity which reaches a maximum value of 1.64 T. The presence of Cu at low concentrations enhances coercivity up to the SmCo$_{3.5}$Cu$_{1.5}$. Following the X-ray diffraction patterns, the 1:5 phase is by far the predominant phase in all samples. A small fraction of the 2:17 phase which seems to be induced from the substrate (rather than to a nanoscale phase decomposition as described previously~\cite{sharma2021epitaxy}, is fully suppressed for higher Cu substitution. XMCD element-specific hysteresis loops proved that Cu substitution results in an increasing decoupling of the Sm and Co moments. Last but not least, electron microscopy measurements showed that the films are dominated by mainly the 1:5 phase. 

We first discuss the effect of Cu substitution on the intrinsic magnetic anisotropy of SmCo$_{5}$, especially the single-ion anisotropy (SIA) of Sm$^{3+}$. While intuitively one would expect an immediate decrease of intrinsic magnetic anisotropy, the computational methods described in Sec.~\ref{comp} show a different and more complex behavior. The calculated CFPs $B_{lm}$ for SmCo$_{5}$, Sm(Co$_{4}$Cu$_{3g}$)$_{5}$ and Sm(Co$_{4}$Cu$_{2c}$)$_{5}$ are listed in Table~\ref{table:Blm}. SmCo$_{5}$ belongs to the point group D$_{6h}$ therefore only $B_{20}$, $B_{40}$, $B_{60}$ and $B_{6\pm6}$ retain. The evaluated $B_{lm}$ values using the Wannier basis are comparable with the measured values. In particular, the magnitude of $B_{20}$ is dominant and determines largely the magnetic anisotropy of Sm$^{3+}$, {\it i.e.}, a negative $B_{20}$ indicates uniaxial magnetic anisotropy. With Cu occupying the $3g$ sites, $B_{20}$ is more negative, indicating that the SIA of Sm$^{3+}$ is indeed enhanced. In contrast, the Cu doping on the $2c$ site tends to lower the SIA of Sm$^{3+}$. Note that the number of nonzero $B_{lm}$ parameters in Sm(Co$_{4}$Cu$_{3g}$)$_{5}$ is larger due to the lower symmetry.

\begin{table}[h!]
	\centering
        \scriptsize
	\caption{\label{table:Blm} Crystal field parameters (in units of Kelvin) for Sm$^{3+}$Co$_{5}$, Sm$^{3+}$(Co$_{4}$Cu$_{3g}$)$_{5}$ and Sm$^{3+}$(Co$_4$Cu$_{2c}$)$_{5}$. For comparison, the experimental values are taken from Ref.~\cite{sankar1975magnetocrystalline}}
	\begin{tabular}{l|c|c|c|c}
		\hline
		\hline
		$B_{lm}$  & SmCo$_{5}$ & Expt. & Sm(Co$_4$Cu$_{3g}$)$_{5}$ & Sm(Co$_4$Cu$_{2c}$)$_{5}$ \\
            (K)  & & & & \\
		\hline
            $B_{20}$ & -1068 & -840 & -1380 & -867\\
            $B_{2\pm2}$  & - & - & 168 & -\\
            $B_{40}$ & 5 & 200 & 77& -91\\
            $B_{4\pm2}$ & - & -& 102 & -\\
            $B_{4\pm4}$ & - & - & -44 & -\\
            $B_{60}$ & -473 & 0 & -472 &-430\\
            $B_{6\pm2}$ & - & -& -14 & -\\
            $B_{6\pm4}$ & - & -& -34 & -\\
            $B_{6\pm6}$ & 494 & 6 & 501 & 456 \\
		\hline
            \hline
	\end{tabular}
\end{table}

\begin{figure}
    \centering
    \includegraphics[width=\columnwidth]{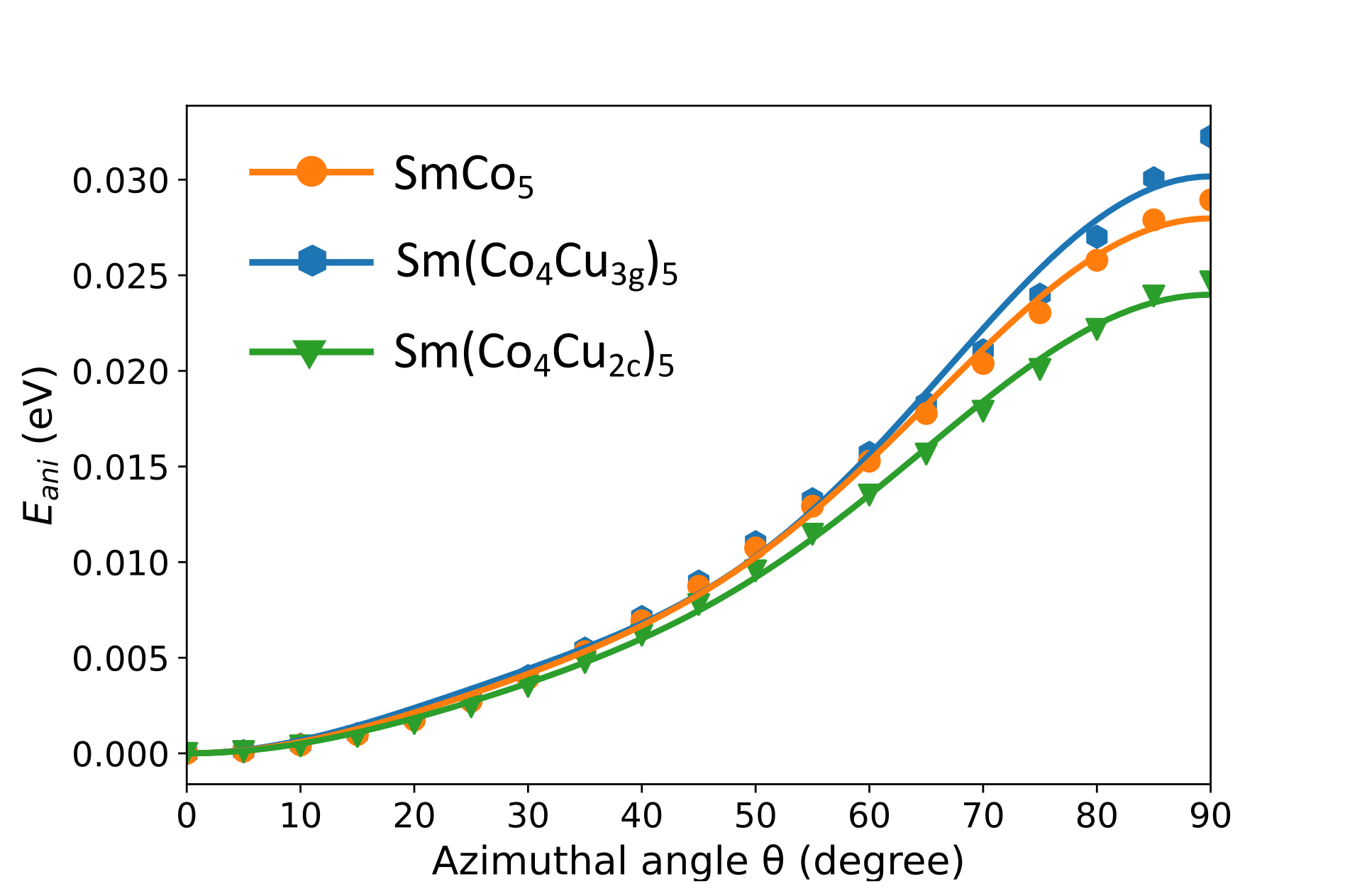}
    \caption{Ground state energy of Sm 4\textit{f} shell in SmCo$_{5}$ (orange circle), Sm(Co$_4$Cu$_{3g}$)$_{5}$ (blue hexagons), Sm(Co$_4$Cu$_{2c}$)$_{5}$ (green triangle) as a function of magnetization direction represented by the azimuthal angle $\theta$.}
    \label{fig:mae}
\end{figure}

Another important aspect to be considered is the exchange coupling between the Sm and its neighboring Co atoms, which can be weakened upon introducing non-magnetic Cu atoms into the system. The exchange field in SmCo$_5$ is set to be 250 T according to previous experimental~\cite{tils1999crystal} and theoretical~\cite{patrick2019temperature} work, while for Sm(Co$_4$Cu$_{3g}$)$_{5}$ and Sm(Co$_4$Cu$_{2c}$)$_{5}$, the exchange fields are rescaled based on the calculated $J_{\text{SmCo}}$ values contributed from the first-nearest neighbors of Sm with respect to the $J_{\text{SmCo}}$ of SmCo$_5$. Such exchange coupling parameters are evaluated using the OpenMX~\cite{terasawa2019efficient} code in the LDA + $U$ regime with $U = 6.7$\,eV and $J = 0.7$\,eV for the Sm 4\textit{f} states. Accordingly, we set the exchange fields of Sm(Co$_4$Cu$_{3g}$)$_{5}$ and Sm(Co$_4$Cu$_{2c}$)$_{5}$ to 201 and 230 T, respectively. The Coulomb interaction parameters and the spin-orbit coupling strength are taken from Ref.~\cite{tripathi2018xmcd}. By varying the exchange field direction which is represented by the azimuthal angle $\theta$, Fig.~\ref{fig:mae} shows the eigenvalue $E_{ani}$ of $\hat{H}_{at}$ as a function of $\theta$. It can be explicitly observed that Sm$^{3+}$ in Sm(Co$_4$Cu$_{3g}$)$_{5}$ possesses the highest SIA while in Sm(Co$_4$Cu$_{2c}$)$_{5}$ the SIA is the lowest. By fitting the energy curve to
\begin{equation}
    E_{ani} (\theta) = K_1 \mathrm{sin}^2\theta + K_2 \mathrm{sin}^4\theta + K_3 \mathrm{sin}^6\theta, 
\end{equation}
we obtain $K_1$ of 21 meV, 24 meV and 17 meV for SmCo$_{5}$, Sm(Co$_4$Cu$_{3g}$)$_{5}$ and Sm(Co$_4$Cu$_{2c}$)$_{5}$, respectively. Note here that we omit the $K_3^{\prime}$ term associated with $K_3^{\prime} \mathrm{sin}^6\theta \mathrm{cos}6\phi$ since we find $K_3^{\prime}$ rather small. Besides, the Cu doping effect on the Co sites can be approximated using YCo$_{5}$ as a prototype. It has been reported that the $3g$ sites doping of Cu increases the magnetic anisotropy of YCo$_{5}$~\cite{okumura2022first} while the $2c$ sites doping is expected to reduce the magnetic anisotropy due to a larger $K_1$ of Co$_{2c}$ than that of Co$_{3g}$. In addition, according to our DFT calculations, the $2c$-site doping is slightly energetically favorable by about 7\,meV/atom as compared to the $3g$-site doping. This strongly implies a statistically random distribution on both sites in reality, as 7 meV amounts to a temperature of 81 K.
Therefore, considering the contrasting roles played by Cu with different Wyckoff position occupations, a non-monotonous change of magnetic anisotropy with Cu doping content can be expected, as also observed in the Ce-Co based systems~\cite{lamichhane2019single,xie2022mixed}. It is intriguing to characterize the intrinsic properties of the Cu-substituted SmCo$_{5}$ system to bridge the gap between the microscopic equilibrium properties and the macroscopic coercivity~\cite{matsumoto2020magnetism}.  

From a macroscopic point of view, Sm-rich precipitates as well as the formation of many different Sm-Co phases and structural disorder phenomena are strongly connected to high coercivity values in sputtered Sm-Co films. \cite{akdogan2015preparation}. Sayama \emph{et al.} \cite{sayama2006origin} suggested that the origin of high perpendicular anisotropy in SmCo$_{5}$ thin films on Cu buffer layers are related to the diffusion of Cu atoms into the SmCo$_{5}$ structure. Note that this diffusion effect is excluded in the presented study. The layer of Sm$_{2}$Co$_{17}$ at the substrate interface will not contribute to a systematic change of the hysteresis itself. As the defect structure is similar throughout the consistent series of thin films, we correlate the increased coercivity to the increased intrinsic magnetocrystalline anisotropy upon Cu substitution and to the inhomogeneous distribution of Cu and Co. Since other phases, grain and phase boundaries can be excluded for our thin film model systems, we suggest the nanoscale disproportionation of Cu and Co as additional novel source of coercivity leading to enhanced pinning with increased Cu substitution.     

\section{Conclusions}
In summary, our investigation of highly crystalline SmCo$_{5-x}$Cu$_{x}$ thin films grown by MBE on Al$_{2}$O$_{3}$ substrates in combination with advanced computational and characterization methods, has revealed that copper substitution enhances the intrinsic magnetic anisotropy, correlated with an increase in coercivity. As a second source of increased coercivity, we have identified with EDX measurements a disproportionation of Cu- and Co-rich areas within an otherwise homogeneous 1:5 phase structure. The practical applicability of Cu to replace Co is limited of course by the induced overall reduction of the total magnetic moment, acting negatively on the ($BH$)$_{\text{max}}$ product. XMCD provided clear evidence of a decoupling of Sm and Co moments in the presence of Cu. Electron microscopy confirmed the presence of small amount of interfacial Sm$_{2}$Co$_{17}$ phase in the highly crystalline SmCo$_{5}$ layer. Based on thin film model systems, our study provides novel insight into the complex materials science and hardening mechanisms in rare-earth-based permanent magnetic materials by disentangling different sources of intrinsic and extrinsic contributions to the hysteresis behavior. 

\begin{acknowledgement}

We acknowledge the financial support from the Deutsche Forschungsge-meinschaft (DFG) in the framework of the CRC/TRR 270 (Project ID. 405553726), projects A02, A03, A05, B05, and Z01/02. JPP acknowledges the German Research Foundation (Deutsche Forschungsgemeinschaft - DFG) for the funding under project 429646908. The authors acknowledge Janghyun Jo for experimental help and Lea Risters for FIB sample preparation. This research used resources of the Advanced Light Source, which is a DOE Office of Science User Facility under contract no. DE-AC02-05CH11231.

\end{acknowledgement}


\bibliography{Gkouzia}

\end{document}